\documentclass[aps,showpacs,prl,twocolumn]{revtex4}

\begin{document}

\noindent{\bf Comment on ``Four-Point Resistance of Individual
Single-Wall Carbon Nanotubes''}
\bigskip

In a recent Letter Gao {\em et al}.[1] have claimed a dramatic
experimental result: the four-probe resistance of a
single carbon nanotube sample can be
significantly negative at low temperature. This appears
to suggest that quantum interference effects from the
voltage probes play a crucial role in
producing this surprising phenomenon. In this comment
we point out a set of serious contradictions.

Gao {\em et al}. have analyzed their results by the multi-terminal
theory of B\"uttiker [2]. We do not comment here on the
Landauer-B\"uttiker method used for a current-driven system
with multiple test probes, with an equal multiplicity of individual
chemical potentials. The invocation of several distinct
chemical potentials, in various parts of the one system, is moot
from the statistical mechanics point of view, and is discussed elsewhere in
the literature [3,4].  

In a four-probe measurement, one attempts to gauge the natural voltage drop
between two separated voltage probes (3 and 4 in this case). Such a procedure
has to be highly noninvasive [5] if a meaningful, intrinsic sample resistance
is to be extracted. ``Noninvasivity'' means no disruption whatsoever
to the current flow inherent in the sample. Therefore, the voltage drop
is a true measure of intrinsic resistance.

Gao {\em et al}. argued that the chemical potentials $\mu_3$ and $\mu_4$
are both
bounded by $\mu_1$ above, and $\mu_2$ below; see Figure 1 of Ref. [1].
The ratio of the four-point and two-point resistances is then
$R_{\rm 4PT}/R_{\rm 2PT} = (\mu_3 - \mu_4)/(\mu_1 - \mu_2)$.
According to B\"uttiker's theory [2], this is equivalent to
$R_{\rm 4PT}/R_{\rm 2PT} = [{\cal T}_{31}{\cal T}_{42}
- {\cal T}_{32}{\cal T}_{41}]/[({\cal T}_{31}+{\cal T}_{32})
({\cal T}_{41}+{\cal T}_{42})]$ in terms of the transmission
probability functions ${\cal T}_{ij}$.
If $\mu_3 > \mu_4$, then $R_{\rm 4PT}$ is positive; but
if $\mu_3 < \mu_4$, then $R_{\rm 4PT}$ is negative.
The transmission functions ${\cal T}_{ij}$ alone
determine the sign.  Gao {\em et al.} seem to believe that
$R_{\rm 4PT} < 0$ comes from strong quantum interference effects. 

To cause significant interference, the ${\cal T}_{ij}$s
must depend sensitively on the
different pathways from the different voltage probes. In the samples of
Ref. [1] the voltage probes are a series of complex multi-wall nanotube
structures. Unlike in the de Picciotto {\em et al}. experiment [5],
here the probes' noninvasiveness is  less transparent in terms of 
measurability and controllability. Unless the voltage probes are {\em known} 
to be noninvasive, the notion of an intrinsic sample resistance has 
little meaning. 
In our view the results and intepretation in Ref. [1]
contain several major contradictions. 

(a) In the analysis of Gao {\em et al}. a negative $R_{\rm 4PT}$ arises 
if and only if the numerator is negative in the
B\"uttiker ratio. For that to happen,
${\cal T}_{32}{\cal T}_{41} > {\cal T}_{31}{\cal T}_{42}$.
A transmission function usually
depends on the height and width of its barrier. For the identical and
symmetrically placed probes 3 and 4, the distances between
3\&2 and 1\&4 exceed 3\&1 and 4\&2. The larger the distance,
the smaller the transmission. Hence,
elementary geometry makes the above condition unlikely.

(b) Now consider the original Landauer 
formulas for two-point and  four-point resistance [3].
$R_{\rm 2PT} = (h/2e^2)/{\cal T}$ and
$R_{\rm 4PT} = (h/2e^2){\cal R}/{\cal T}$, where
${\cal T}$ and ${\cal R}$ are total transmission and reflection functions
respectively. The formula for $R_{\rm 4PT}$ can be derived from the
B\"uttiker relation; see Eq. (3.301) of Ferry and Goodnick [6].
The eigenvalues of ${\cal R}$ and ${\cal T}$ are bounded by 0 and 1,
and ${\cal R} = 1 - {\cal T} > 0$ must conserve unitary probability.

Because  $R_{\rm 2PT}$ is positive, ${\cal T}$ is certainly positive. 
But if $R_{\rm 4PT}$ is negative, as in Gao {\em et al}., it means
that ${\cal R} < 0$. Therefore ${\cal T} > 1$.
How does this conserve probability?

(c) $R_{\rm 4PT}$ is the desired sample resistance.
It is a linear-response quantity measured at small applied fields.
At zero temperature, it is an {\em inherent} property of
the equilibrium ground state for the (presumably thermodynamically stable)
device. If a well calibrated noninvasive method were to be used, one would
expect $R_{\rm 4PT}$ to be zero (ideal metal/ballistic case) [5]
or very small, depending on any residual dissipation.
The intrinsic resistance of the sample must, at the very least,
be nonnegative to respect the Joule heating law $W = IV \ge 0$
and thus to guarantee thermodynamic stability.

We wish to thank Dr Adrian Bachtold for useful correspondence.

\bigskip
[1] B. Gao {\em et al}. Phys. Rev. Lett. {\bf 95}, 196802 (2005).

[2] M. B\"uttiker, IBM J. Res. and Dev. {\bf 32}, 317 (1988).

[3] A. Kamenev and W. Kohn, Phys. Rev. B {\bf 63}, 155304 (2001).

[4] M. P. Das and F. Green, J. Phys: Condens.Matter {\bf 15}, L687 (2003).

[5] R. de Picciotto {\em et al}.,  Nature (London) {\bf 411}, 51 (2001).

[6] D. K. Ferry and S. M. Goodnick, {\em Transport in Nanostructures}         
    (Cambridge Univ. Press, Cambridge, 1997).

\bigskip
M.P. Das${}^1$, F. Green${}^1$, and J. S. Thakur${}^2$

${}^1$ Dept. of Theoretical Physics, RSPhysSE,
The Australian National University, Canberra, ACT 0200, Australia.

${}^2$  Dept. of Electrical and Computer Eng., Wayne State University,
Detroit, Michigan, 48202, USA.

\end{document}